\documentclass
[twocolumn,a4paper,superscriptaddress,floatfix,showpacs,aps]{revtex4}%
\usepackage[english]{babel}
\usepackage[utf8]{inputenc}
\usepackage{amsmath}
\usepackage{graphicx,epstopdf}
\usepackage{amssymb,epsfig,color,textcase}
\usepackage{hyperref}
\usepackage{amsfonts}
\usepackage{amssymb}
\usepackage{ulem}
\usepackage{graphicx}%
\providecommand{\U}[1]{\protect\rule{.1in}{.1in}}

\begin{document}
\title{Spectroscopic signatures of molecular orbitals on a honeycomb lattice.}
\author{Z.V. Pchelkina}
\email{pzv@ifmlrs.uran.ru}
\affiliation{M.N. Miheev Institute of Metal Physics of Ural Branch of Russian Academy of
Sciences, 620137, Ekaterinburg, Russia}
\affiliation{Theoretical Physics and Applied Mathematics Department, Ural Federal
University, Mira St. 19, 620002 Ekaterinburg, Russia}
\author{S. V. Streltsov}
\affiliation{M.N. Miheev Institute of Metal Physics of Ural Branch of Russian Academy of
Sciences, 620137, Ekaterinburg, Russia}
\affiliation{Theoretical Physics and Applied Mathematics Department, Ural Federal
University, Mira St. 19, 620002 Ekaterinburg, Russia}
\author{I. I. Mazin}
\affiliation{Code 6393, Naval Research Laboratory, Washington, DC 20375, USA}
\date{\today}

\begin{abstract}
A tendency to form benzene-like molecular orbitals has been recently shown to
be a common feature of the $4d$ and $5d$ transition metal oxides with a
honeycomb lattice. This tendency competes with other interactions such as the
spin-orbit coupling and Hubbard correlations, and can be partially or
completely suppressed. In the calculations, SrRu$_{2}$O$_{6}$ presents the
cleanest, so far, case of well-formed molecular orbitals, however, direct
experimental evidence for or against this proposition has been missing. In
this paper, we show that combined photoemission and optical studies can be
used to identify molecular orbitals in SrRu$_{2}$O$_{6}$. Symmetry-driven
election selection rules suppress optical transitions between certain
molecular orbitals, while photoemission and inverse photoemission measurements are insensitive to them. Comparing the photoemission and optical conductivity spectra one should be able to observe  clear signatures of molecular orbitals.

\end{abstract}
\maketitle

\textit{Introduction.} Low dimensional ruthenates with a honeycomb lattice
have been attracting a lot of attention in recent years. $\alpha-$RuCl$_{3},$
which has one hole in the $t_{2g}$ manifold, shows hallmarks of Kitaev
physics\cite{Plumb2014,Sears2015}, Li$_{2}$RuO$_{3}$ with two $t_{2g}$ holes
dimerizes in the low-temperature phase\cite{Miura2007,Wang2014} and exhibits a
valence bond liquid behavior at high temperatures~\cite{Kimber2014,Park2016},
while SrRu$_{2}$O$_{6}$ with a half-filled $t_{2g}$ band shows rather unusual
magnetic properties\cite{Hiley2014}. It has been argued~\cite{Streltsov2015}
that the physics of these compounds is underscored by competition between the
spin-orbit coupling and Hubbard correlations, on one side, direct Ru-Ru
one-electron hopping, on the other side, and O-assisted indirect hopping that
leads to formation of molecular orbitals (MO), on the third
side~\cite{Kim2016}. \textit{Ab initio} calculations show that MOs appear to
dominate in the last compound~\cite{Streltsov2015}. In the first two they are
mostly suppressed, but at least in $\alpha-$RuCl$_{3}$ (and in a similar
compound, Na$_{2}$IrO$_{3})$ they manifest themselves $via$ an anomalously
large third-neighbor coupling~\cite{Winter}.

MOs inevitably occur if transition metals with active $t_{2g}$ orbitals form a
honeycomb lattice and $t_{2g}$ electrons can only hop via oxygen $p$ orbitals
\cite{Mazin2012}. In this case, the electronic structure problem maps onto
that of the benzene molecule, essentially, a 6-member ring with nearest and
next-nearest neighbor hoppings only ($t_{1}^{\prime}$ and $t_{2}^{\prime},$
respectively). The electronic structure then consist of four levels, $A_{1g}$,
$E_{1u}$, $E_{2g}$, $B_{1u}$ ($E_{1u}$ and $E_{2g}$ are doubly degenerate),
formed by six molecular orbitals. Their energies are: $E_{A_{1g}}%
=2(t_{1}^{\prime}+t_{2}^{\prime})$, $E_{E_{1u}}=(t_{1}^{\prime}-t_{2}^{\prime
})$, $E_{E_{2g}}=-(t_{1}^{\prime}+t_{2}^{\prime})$, and $E_{B_{1u}}%
=-2(t_{1}^{\prime}-t_{2}^{\prime})$~\cite{corr}. In this approximation, an
electron occupying one of the MOs remains fully localized within one of the Ru
hexagons, in spite of the fact that the lattice itself is uniform without any
dimerization or clusterization~\cite{Foyevtsova2013}.

In real materials $t_{1}^{\prime}/3\sim-t_{2}^{\prime}>0$ and the two highest
MO levels, $A_{1g}$ and $E_{1u}$, turn out to be nearly degenerate
\cite{Foyevtsova2013,Streltsov2015}. This is conducive for the spin-orbit
coupling (SOC) and is the reason why the SOC is so efficient in the case of
one $t_{2g}$ hole as in $\alpha-$RuCl$_{3}$ or in Na$_{2}$IrO$_{3}$. Moreover,
for the whole range between the weak and the strong SOC limit the highest
energy state ($j_{eff}=1/2$ or $A_{1g}$ in the respective limits) is
half-filled and therefore Hubbard correlations are important.

Increasing number of holes, i.e. going from Ru$^{3+}$ to Ru$^{4+}$ makes
$E_{1u}$ band half-filled. One may lift the degeneracy and gain some energy
not due to the SOC or formation of molecular orbitals on hexagons, but
dimerizing lattice (if the elastic energy penalty would not be too large). In
this case the system gains considerable covalent energy due to direct $d-d$
hopping (which may be large in the common edge geometry) and forms
spin-singlet dimers. This scenario is realized in Li$_{2}$RuO$_{3}%
$\cite{Miura2007,Kimber2014}.

In the case of three $t_{2g}$ holes, Ru$^{5+}$, we arrive at the situation,
when $A_{1g}$ and $E_{1u}$ states are completely empty and the MOs with their
large gap between the $E_{1u}$ and $E_{2g}$ states are energetically
favorable. In the ionic approximation the energy gain is of the order of
$E_{E_{1u}}-E_{E_{2g}}\approx2t_{1}^{\prime}$. Interestingly, the long range
Ne\'{e}l antiferromagnetic (AFM) order does not destroy MOs, but even
increases this energy gain\cite{Streltsov2015}. These are the reasons why the
MOs are so clearly seen in the band structure calculations in SrRu$_{2}$%
O$_{6}$\cite{Streltsov2015}.

While MO scenario has been very successful in explaining the physical
properties of SrRu$_{2}$O$_{6}$\cite{Streltsov2015}, no direct observation of
MOs has been effected so far, and other, albeit, in our opinion, less
convincing, scenarios have been proposed\cite{Tian2015}. In this paper we suggest that a combination of the spectroscopic techniques sensitive and insensitive to the dipole selection rules may provide direct evidence of the formation of MOs in SrRu$_{2}$O$_{6}.$ These can be, \textit{e.g.}, optical absorption and photoemission measurements (the latter are mostly determined by the electronic density of states, DOS), properly corrected for corresponding
cross sections. We will show both analytically and numerically that the
optical conductivity in the MO picture is dramatically different from the
joint DOS, because of unusually restrictive optical selection rules.
\begin{figure}[t]
\includegraphics[width=0.9\columnwidth,angle=0]{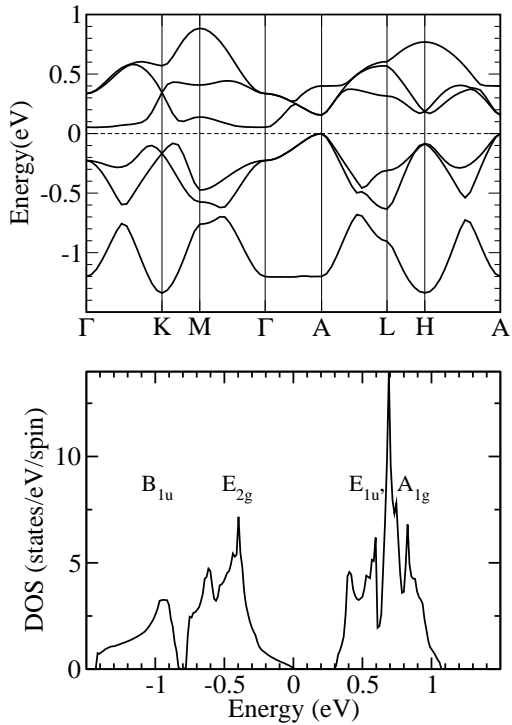}\caption{The
nonmagnetic GGA band structure (upper panel) and total DOS obtained by GGA
calculation for the Ne\'{e}l antiferromagnetic structure (lower panel). The
contributions from different molecular orbitals are labeled according to
Refs.~\cite{corr, Streltsov2015}.}%
\label{dos_bands}%
\end{figure}

\textit{Optical properties of molecular orbitals.} The dipole selection rules
prohibit optical transitions between states of the same parity. In the MO picture, this leaves four transitions: $B_{1u}\rightarrow E_{2g}$ (at
$\hbar\omega=t_{1}^{\prime}-3t_{2}^{\prime})$, $E_{2g}\rightarrow E_{1u}$ (at
$2t_{1}^{\prime})$, $E_{1u}\rightarrow$ $A_{1g}$ (at $t_{1}^{\prime}%
+3t_{2}^{\prime})$, and $B_{1u}\rightarrow A_{1g}$ (at $4t_{1}^{\prime}).$ For
the half filling, representative of SrRu$_{2}$O$_{6}$, that would generate two
absorption peaks, corresponding to the $E_{2g}\rightarrow E_{1u}$ and
$B_{1u}\rightarrow A_{1g}$ transitions, the latter at a twice larger energy
than the former. However, there is an additional symmetry in the problem that
forbids some of these transitions. Indeed, to assure a nonzero optical matrix
element, the direct product of the representations of the initial and final
states must contain a representation of the corresponding component of the
dipole operator $p^{\alpha}$ (see, \textit{e.g.}, Ref.~\cite{Flurry}). In the
case of an ideal hexagon with the point group symmetry $D_{6h}$ the $p^{x}$
and $p^{y}$ components are transformed according to the $E_{1u}$
representation~\cite{Wilson1955}. Since
\begin{align}
B_{1u}\times A_{1g} &  =B_{1u}\\
B_{1u}\times E_{2g} &  =E_{1u}\\
E_{2g}\times E_{1u} &  =B_{1u}+B_{2u}+E_{1u}\\
E_{1u}\times A_{1g} &  =E_{1u}%
\end{align}
the point symmetry will suppress $B_{1u}\rightarrow A_{1g}$, but not
$B_{1u}\rightarrow E_{2g}$, $E_{2g}\rightarrow E_{1u}$, and $E_{1u} \to
A_{1g}$ transitions. In SrRu$_{2}$O$_{6}$ only $E_{2g}\rightarrow E_{1u}$
transitions are allowed, but in other hexagonal systems with different number
of $d$ electrons one may also expect $B_{1u} \to E_{2g}$ and
$E_{1u}\to A_{1g}$ transitions. In the Appendix we show explicitly the
matrix elements of $p^{\alpha}$ in the nearest- and next-nearest neighbor
tight binding approximation. The out-of plane matrix element is zero and
corresponding optical transitions are absent in the MO approximation.

Together with the selection rules forbidding transitions between states with
the same parity this additional selectivity offers a direct test of the MO
scenario. It suggests that despite the double-hump structure of the DOS
(Fig.~\ref{dos_bands}), and, correspondingly, joint DOS, the optical
absorption $\sigma(\omega)$ will have a one peak structure. Importantly, this
is a qualitative, not quantitative test. While the exact positions and
relative intensities of different peaks in DOS and $\sigma(\omega)$ may differ
from the density function theory predictions (due to many-body effects), the
general structure described above should qualitatively hold. This way one can
directly verify by spectroscopical means (comparing optical, photoemission and
inverse photoemission spectra) the concept of molecular orbitals.
\begin{figure}[t]
\begin{center}
\includegraphics[width=1.0\columnwidth,angle=-90]{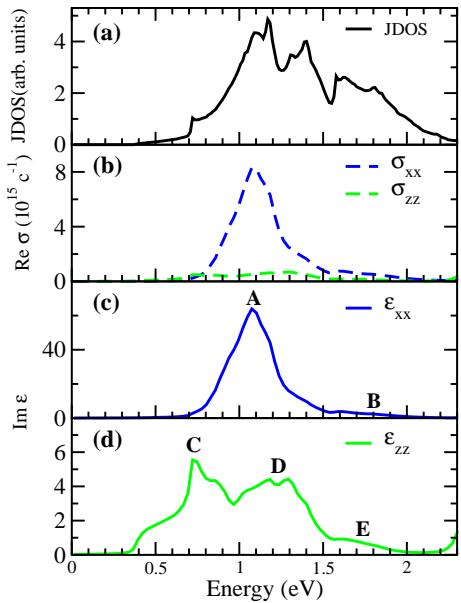}
\end{center}
\caption{Results of the antiferromagnetic GGA calculations. (a) The joint
density of states, $J(\omega)$, is shown by black line. (b) The real part of
optical conductivity, $\mathrm{Re}~\sigma_{\alpha\beta}(\omega)=\frac{\omega
}{4\pi}\mathrm{Im}~\varepsilon_{\alpha\beta}(\omega)$, where $\alpha, \beta=
x$ (blue dotted line) and $\alpha, \beta= z$ (green dotted line). The
imaginary part of the frequency-dependent dielectric functions, $\varepsilon
_{xx}$ (c) and $\varepsilon_{zz}$ (d) are shown by solid blue and green lines,
correspondingly.}%
\label{im_epsilon}%
\end{figure}

\textit{DFT calculations of }$\sigma(\omega)$ \textit{in SrRu}$_{2}$%
\textit{O}$_{6}$\textit{.} We used the full-potential linearized augmented
plane-wave (LAPW) method as implemented in the WIEN2k code~\cite{wien2k} to
calculate optical properties of SrRu$_{2}$O$_{6}$. We used the
exchange-correlation potential of Ref.~\cite{pbe}. Integration was performed
using the tetrahedron method on a mesh consisting of 4096 k-points in the
Brillouin zone (BZ). The radii of atomic spheres were chosen to be 2.36, 1.93
and 1.72 a.u. for Sr, Ru, and O, respectively. The parameter of the plane wave
expansion was set to $R_{MT}K_{max}$=7, where $R_{MT}$ is the radius of O and
$K_{max}$ is the plane wave cut-off.

For a dielectric, the imaginary part of the dielectric function $\operatorname{Im}%
\varepsilon(\omega)=4\pi\sigma(\omega)/\omega$ in the random-phase  approximation (RPA) is defined as
\begin{align}
\mathrm{Im}~\varepsilon_{\alpha\beta}(\omega) &  =\frac{e^{2}}%
{\pi m^{2}\omega^{2}}\sum_{c,v}\int\langle c,\mathbf{k}|p^{\alpha}|v,\mathbf{k}%
\rangle\langle v,\mathbf{k}|p^{\beta}|c,\mathbf{k}\rangle
\nonumber\label{mainEq}\\
&  \times\delta(\epsilon_{c}(\mathbf{k})-\epsilon_{v}(\mathbf{k})-\hbar
\omega)d\mathbf{k}.
\end{align}
where $m$ is the electron mass, $\{\alpha,\beta\}=\{x,y,z\}$, summation runs
over all pairs of conduction (c) and valence (v) bands, and $\epsilon
(\mathbf{k})$ gives the energy of corresponding band, while $\langle
c,\mathbf{k}|p^{\alpha}|v,\mathbf{k}\rangle$ is the momentum operator's matrix
element~\cite{Ambrosch2006}. This, obviously, includes the phase space factor,
usually called the joint density of states,
\[
J(\omega)=\sum_{c,v}\int\delta(\epsilon_{c}(\mathbf{k})-\epsilon
_{v}(\mathbf{k})-\hbar\omega)d\mathbf{k},
\]
and the effects of the matrix elements. The $J(\omega)$ obtained within the
AFM GGA calculations is shown in Fig.~\ref{im_epsilon}(a). One observes a
broad maximum in the joint DOS at 1.1--1.4 eV, due to the transitions between
the $E_{2g}$ and the $E_{1u}+A_{1g}$ manifolds, and another maximum at 1.6-1.8
eV, due to the $B_{1u}\rightarrow E_{1u}+A_{1g}$ transitions.

Since SrRu$_{2}$O$_{6}$ has a trigonal crystal structure there are only two
independent components in the dielectric tensor,
$\varepsilon_{xx}$ and $\varepsilon_{zz}$. Fig.~\ref{im_epsilon}(c), (d) shows
the calculated imaginary part of dielectric tensor components for SrRu$_{2}%
$O$_{6}$. The amplitude of the $\varepsilon_{xx}$ component is about 8 times
larger than the one of $\varepsilon_{zz}$, reflecting the fact that it only
appears through deviations from the MO model. More interestingly, we observe
that $\operatorname{Im}\varepsilon_{xx}(\omega)$ has one strong peak
\textquotedblleft A\textquotedblright at $\sim$1 eV, corresponding to
$E_{2g}\rightarrow E_{1u}$ transitions, while the second peak of $J(\omega)$
is completely suppressed in $\operatorname{Im}\varepsilon_{xx}(\omega)$ (Fig.
~\ref{im_epsilon}(c)). Moreover, the first peak also becomes sharper,
reflecting the fact that, while the $E_{1u}$ and $A_{1g}$ orbitals are
strongly mixed, the higher energy part of the corresponding manyfold has
somewhat more of the $A_{1g}$ character, leaving less room for the
$E_{2g}\rightarrow E_{1u}$ transitions (remember that the $E_{2g}\rightarrow
A_{1g}$ transitions are forbidden by parity). This is exactly the qualitative
effect we were looking for.

Note that if the matrix elements in Eq. (\ref{mainEq}) are set to a constant,
$\langle c,\mathbf{k}|p^{\alpha}|v,\mathbf{k}\rangle=const$, then
$\omega\sigma(\omega)=const\cdot J(\omega),$ and, indeed often in
computational papers joint DOS is compared to $\omega\sigma(\omega)$. However,
in real materials, $|\langle c|\mathbf{p}|v\rangle|^{2}/m$ usually grows with
energy, roughly as $(E_{c}-E_{v})$~\cite{footnote}, so one can elucidate the
suppression of particular transitions by comparing $J(\omega)$
(Fig.~\ref{im_epsilon}(a)) with $\sigma(\omega)$ (Fig.~\ref{im_epsilon}(b)).

It is worth noting that the structure of $\operatorname{Im}\varepsilon
_{zz}(\omega) $, which cannot be derived from the MO model, is nonetheless
quite interesting. Indeed, the $p^{z}$ matrix element appears to be strongly
enhanced in the very low frequency region, from the absorption edge to about
0.7 eV (the feature denoted \textquotedblleft C\textquotedblright\ in Fig.
~\ref{im_epsilon}(d)). The matrix elements for next feature, \textquotedblleft
D\textquotedblright, are suppressed by a factor of $\approx1.5$ [2.2-2.3 in
$\operatorname{Im}\varepsilon_{zz}(\omega)/J(\omega)],$ and the high-energy
region corresponding to the $B_{1u}\rightarrow E_{1u}+A_{1g}$ transitions by
an additional factor of $\approx$1.8 (feature \textquotedblleft
E\textquotedblright).

Compared to iridates Na$_{2}$IrO$_{3}$ and Li$_{2}$IrO$_{3},$ often quoted in
the context of MOs, SrRu$_{2}$O$_{6}$ has a clear advantage in the sense that
in iridates the MO picture is contaminated by a strong spin-orbit interaction
that makes selection rules not well expressed. Indeed, while DFT
calculations~for iridates~\cite{Li2015} agree well with experimental data, they
cannot be interpreted in such a simple way as ours presented above, and cannot
provide such a qualitative assessment of the MO picture.

\textit{Conclusions.} We presented first principle calculations of the optical
properties of the putative molecular orbital solid SrRu$_{2}$O$_{6}$, as well
as an analytical analysis of the optical absorption in the molecular orbitals
model. We have identified a qualitative signature of molecular orbitals in
optical properties. There are only four possible transitions allowed by the
parity of the wave functions, but one of these parity-respecting optical
transitions is suppressed by the point group symmetry, an unusual effect
directly related to molecular orbitals. Different distortions of the crystal
lattice, spin-orbit coupling, correlation effects etc. may completely suppress
formation of molecular orbitals or strongly modify their structure. Our
results show that one may use optical spectroscopy as a probe to study
molecular orbital physics in transition metals oxides consisting of honeycomb layers.

\textit{Acknowledgements.} We are grateful to V. Anisimov and R. Valent{\'i} for useful discussions. This work was supported by Civil Research and Development Foundation via program FSCX-14-61025-0, the Russian Foundation of Basic Research via Grants No. 16-02-00451. SVS and ZVP were additionally supported by FASO (theme \textquotedblleft Electron\textquotedblright\ No. 01201463326) and Russian ministry of education and science via act 11 contract 02.A03.21.0006, while IIM was supported by ONR through the NRL basic research program.

\textit{Appendix: tight-binding treatment of optical properties in an ideal MO system.} While there are three $t_{2g}$ orbitals on each Ru site, so that
formally the tight-binding (TB) Hamiltonian is 18$\times$18, only one $t_{2g}$
orbital per site contributes to any given MO\cite{Mazin2012}, so the problem
is reduced to 6$\times$6. This allows us to map the full $t_{2g}$ problem onto
a simple tight-binding model on an ideal hexagon with one $s$-orbital per
site:%
\begin{eqnarray}
\label{hamilt}
H=%
\begin{pmatrix}
0 & t_{1}^{\prime} & t_{2}^{\prime} & 0 & t_{2}^{\prime} & t_{1}^{\prime}\\
t_{1}^{\prime} & 0 & t_{1}^{\prime} & t_{2}^{\prime} & 0 & t_{2}^{\prime}\\
t_{2}^{\prime} & t_{1}^{\prime} & 0 & t_{1}^{\prime} & t_{2}^{\prime} & 0\\
0 & t_{2}^{\prime} & t_{1}^{\prime} & 0 & t_{1}^{\prime} & t_{2}^{\prime}\\
t_{2}^{\prime} & 0 & t_{2}^{\prime} & t_{1}^{\prime} & 0 & t_{1}^{\prime}\\
t_{1}^{\prime} & t_{2}^{\prime} & 0 & t_{2}^{\prime} & t_{1}^{\prime} & 0
\end{pmatrix}
,
\end{eqnarray}
where $t_{1}^{\prime}$ and $t_{2}^{\prime}$ are the nearest and next-nearest neighbor hoppings via oxygen. Diagonalization of this Hamiltonian gives the spectrum described in the introduction.

The dielectric function $\mathrm{Im}~\varepsilon_{\alpha\beta}(\omega)$ in Eq. \eqref{mainEq} is determined by matrix elements of momentum operator $\langle c,\mathbf{k}|p^{\alpha}|v,\mathbf{k}\rangle$, which can be easily calculated using the matrix elements of the momentum operator in the initial TB basis of $s-$orbitals, defined as~\cite{graf}
\[
\mathbf{p}_{ij}=\frac{im}{\hbar}H_{ij}(\mathbf{R}_{i}-\mathbf{R}_{j}),
\]
where $\mathbf{R}_{i}$ and $\mathbf{R}_{j}$ are corresponding sites in the hexagon.

The optical transitions can be characterized by their oscillator strengths
\[
f_{cv}=\frac{2}{m}\frac{\lvert\langle c,\mathbf{k}|p^{\alpha
}|v,\mathbf{k}\rangle\rvert^{2}}{E_{c}-E_{v}},
\]
which can be calculated in the basis of the MOs using eigenvectors of Eq.~\eqref{hamilt} as a transformation matrix. In our model there are only three nonzero momentum operator matrix elements for arbitrary filling of the $d$ shell
\begin{align}
f_{B_{1u},E_{2g}} &  =\frac{ma^{2}}{2\hbar^2}(t_{1}^{\prime}-3t_{2}^{\prime
}),\\
f_{E_{2g},E_{1u}} &  =\frac{ma^{2}}{\hbar^2}t_{1}^{\prime},\\
f_{E_{1u},A_{1g}} &  =\frac{ma^{2}}{2\hbar^2}(t_{1}^{\prime}+3t_{2}^{\prime
}),
\end{align}
where $a$ is the distance between the nearest neighbors (3.0053 \AA~in SrRu$_2$O$_6$). This
is in agreement with symmetry consideration presented above and results to a
single optical $E_{2g}\rightarrow E_{1u}$ transition in SrRu$_{2}$O$_{6}$.

For other fillings, {\it e.g.}, four or two electrons per transition metal site, one
may expect two other transitions, which can be, however, suppressed not due to
the symmetry or parity reasons, but because of a particular ratio between
hopping parameters. \textit{E.g.}, in both RuCl$_{3}$ and SrRu$_{2}$O$_{6},$ as
well as in Na$_{2}$IrO$_{3},$ the hopping $t_{2}^{\prime}$ was found to be of
order of $-t_{1}^{\prime}/3$~\cite{Mazin2012,Wang2015}, which will result in a
strong suppression of the $E_{1u}\rightarrow A_{1g}$ transition. If one chose
$t_{1}^{\prime}=0.3$ eV and $t_{2}^{\prime}=-0.1$ eV as it was estimated for
SrRu$_{2}$O$_{6}$ by Wang \textit{et al.}~\cite{Wang2015}, then indeed
$f_{E_{1u},A_{1g}}\sim0$, while $f_{B_{1u},E_{2g}}$=$f_{E_{2g},E_{1u}}$=0.356.
\begin{figure}[t!]
\begin{center}
\includegraphics[width=0.7\columnwidth,angle=-90]{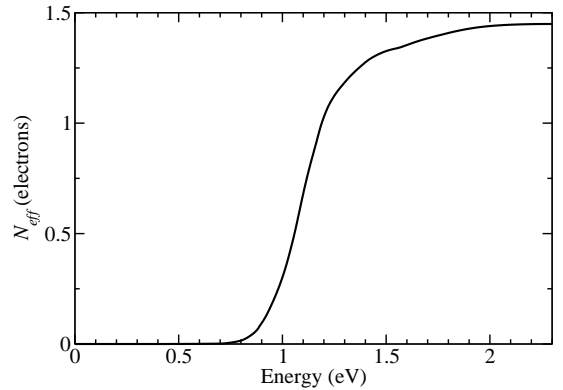}\caption{The effective number of electrons obtained for SrRu$_{2}$O$_{6}$ in the GGA calculation according to Eq. (\ref{n_eff}) for $\varepsilon_{xx}$ component.}%
\label{neff}%
\end{center}
\end{figure}

This provides us with an interesting quantitative check of the validity of
the MO model as regards to the full all-electron DFT calculations. A major
integral characteristic of the optical absorption is given by the optical sum
rule, conveniently written in terms of the effective number of electrons:%
\begin{eqnarray}\label{n_eff}
\int\limits_{o}^{\omega}Im~\varepsilon_{\alpha\alpha}(\omega\prime)\omega^{\prime
}d\omega\prime=\frac{2\pi^{2}e^{2}}{m\Omega}N_{eff}(\omega),
\end{eqnarray}
where $\Omega$ is the unit cell volume. 

The $N_{eff}$ obtained within \textit{ab initio} calculation from 
$xx$ component of the dielectric function for SrRu$_{2}$O$_{6}$ is shown in Fig.~\ref{neff}. A plateau in $N_{eff}(\omega)$ curve clearly points to a presence of a single transition in agreement with model and symmetry considerations. For the energy of 2 eV $N_{eff}^{xx}$=1.44~\cite{sumrule}. In the MO model there is one allowed transition, $E_{2g}\rightarrow E_{1u},$ $f=0.356$ (using the parameters presented above), and, accounting for symmetry and spin degeneracies, {$N_{eff}^{model}=4f=1.424,$ in excellent agreement with the DFT calculations.


\begin{thebibliography}{99}
\bibitem {footnote}One can understand that from the fact that this relation is
exact for the plane waves.
\bibitem {sumrule}There are three valence electrons in SrRu$_{2}$O$_{6},$ of
which this sum rule is missing about 50\%; this is mainly due to the missing
contribution from the $t_{2g}-e_{g}$ transitions. It is well known that this
geometry induces a large O-assisted nearest-neighbor $t_{2g}-e_{g}$ hopping
(proportional to the product $t_{pd\pi}t_{pd\sigma},$ as opposed to the
$t_{pd\pi}^{2},$ as is the case for $t_{2g}-t_{2g}$ $t'_1$).
\end{thebibliography}
\end{document}